\newif\ifabstract
\abstracttrue
 \abstractfalse 
\newif\iffull
\ifabstract \fullfalse \else \fulltrue \fi

\documentclass[11pt]{article}
\usepackage{amsfonts}
\usepackage{amssymb}
\usepackage{amstext}
\usepackage{amsmath}
\usepackage{xspace}
\usepackage{theorem}
\usepackage{graphicx}
\usepackage{url}
\usepackage{graphics}
\usepackage{colordvi}
\usepackage{colordvi}
\usepackage{subfigure}
\usepackage{tabularx}
\usepackage{booktabs} 

\advance \oddsidemargin by -0.8in
\newcommand{\myparskip}{3pt}
\parskip \myparskip

\begin{document}

\title{Nonintrusive Load Monitoring for Machines used in Manufacturing}

\author{Christian Gebbe\thanks{Fraunhofer Research Institution for Casting, Composite and Processing Technology (IGCV), Provinostraße 52, 86150 Augsburg, Germany. Email: {\tt christian.gebbe@gmail.com}}}

\begin{titlepage}
\maketitle

\thispagestyle{empty}

\begin{abstract}
In order to increase the electric energy efficiency of production machines, it is necessary to determine the energy demand of the constituent electric loads. Therefore, a new measurement system based on nonintrusive load monitoring is proposed in this paper. It only measures the voltage and current of the aggregate load and then uses automatic disaggregation methods to estimate the energy demand of the constituent loads. In two case studies, the energy demand of most loads could be determined with an accuracy of 85~\% or more in this way.
\end{abstract}


\end{titlepage}

\section{Introduction}\label{sec: intro}
Increasing the energy efficiency has become a relevant objective for many manufacturing companies \cite{abele2011zukunft, putz2017fraunhofer}. For example, the car manufacturing companies BMW AG, Daimler AG, Volkswagen AG and General Motors Company LLC proclaimed to reduce their energy consumption in the production phase by 20~\% to 45~\% per vehicle \cite{bmw2016sustainable, daimler2016sustainability, volkswagen2016sustainability, gm2016sustainability}.

One of the ways a manufacturing company can increase the energy efficiency is by increasing the electric energy efficiency of their production machines. This can be achieved by e. g. putting machine components into a standby mode in idle times \cite{li2011investigation,Reinhart2016}, modifying process parameters \cite{wang2017dynamic, oda2012study, rief2017evaluation} or by substituting components with more efficient ones \cite{abele2013schlussbericht, bohner2014derivation}.

In order to identify and evaluate such efficiency measures, measuring the electric energy demand of the machine components, i.e. not only the machine as a whole, is recommended by several systematic approaches for increasing the electric energy efficiency of machines \cite{bohner2014derivation, dorr2013methodology, liebl2018approach, reinhart2010energiewertstrom, gotze2012integrated}. How to perform such a measurement is specified further by several authors \cite{abele2015data, kara2011electricity, mohammadi2017methodology, behrendt2012development, thiede2013sme, rao2017conducting, bogdanski2012energy}.

The typically applied measurement method is to acquire the current and voltage for each load. To this end, current clamps need to be installed at each electric load and at each of the three phases in the cabinet box of a machine. This measurement method exhibits high initial and running costs \cite{kara2011electricity, o2013industrial}: First, the initial costs for dozens of current clamps and a data acquisition device which can record dozens of channels synchronously quickly exceed 10,000 or 20,000~EUR. Second, the current clamps can only be installed by a certified electrician by German law. Third, the electrician usually needs to study the wiring diagram of the machine at first. Fourth, there is often little space around the cables which impedes the installation of the current clamps. Fifth, the data acquisition device needs to be configured for each measurement. Sixth, the measurements need to be manually validated in order to prevent common errors such as permutated phases, disconnected sensors or a deviation of the actual wiring from the specifications in the wiring diagram. Thus, setting up a measurement can consume several staff hours up to a day considering all involved persons. Because of these high costs a more economic measurement method is sought after.

\begin{figure} [htb!]
	\centering
	\includegraphics[width=0.8\linewidth]{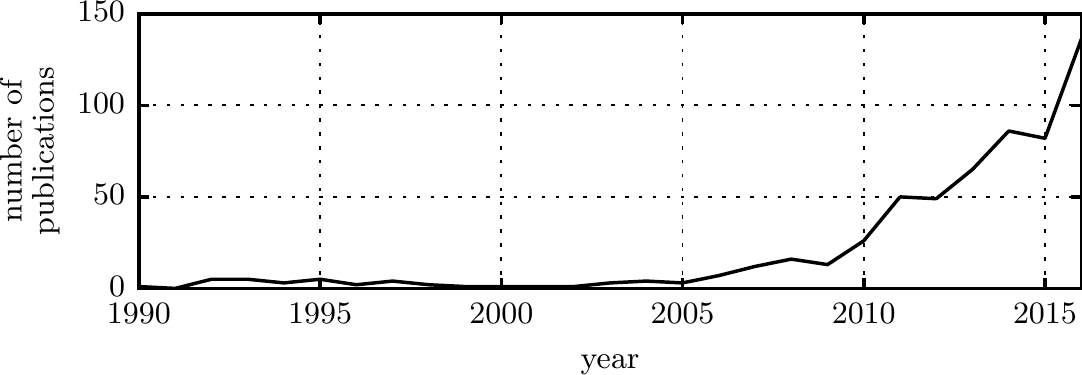}
	\caption{Number of publications per year with the topic nonintrusive load monitoring (in different spellings) according to scopus}
	\label{fig:01_num_publications}
\end{figure}

One possible alternative measurement method is nonintrusive load monitoring. This term refers to methods that estimate the power consumption and operational schedule of electric loads based on only a detailed analysis of the measured current and the measured voltage of the aggregate load \cite{Hart1985, Hart1992, Berges2010, Anderson2012}. In other definitions, more information such as the operating status of the electrical loads or additional sensors can be used as auxiliary input for performing the disaggregation. Despite the increasing popularity of nonintrusive load monitoring (see Fig.~\ref{fig:01_num_publications}), it has so far primarily been used to determine the energy demand of household appliances, but never to determine the energy demand of components of production machines. The closest publication known to the author is \cite{Panten2016}, in which the control signals of the machine components need to be measured to be able to perform the disaggregation. Because of this necessary additional input, the method explained in \cite{Panten2016} does not match the definition of nonintrusive load monitoring used in this paper.

The rest of the paper is structured as follows: First, the new measurement system based on nonintrusive load monitoring is briefly described in section 2. The core of the system consists of multiple algorithms for disaggregation and labeling, which are explained in detail in section 3 and 4, respectively. Section 5 comprises two case studies, in which the measurement system was used to estimate the electric energy demands of all constituent loads of a thermoform machine and a milling machine. Finally, the paper is summarized in section 6.

\section{Measurement system}

\begin{figure} [htb!]
	\centering
	\includegraphics[width=0.8\linewidth]{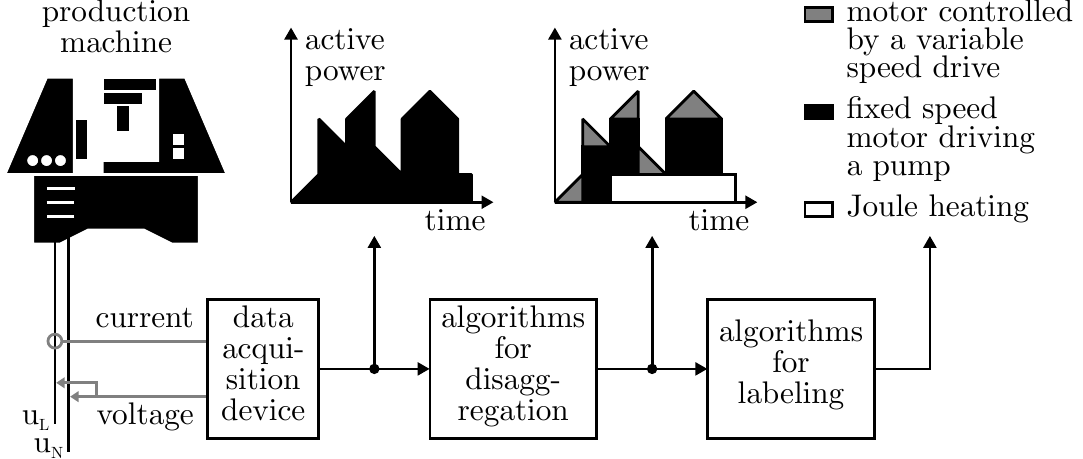}
	\caption{Overview of the proposed measurement system}
	\label{fig:02_system_overview}
\end{figure}

The main part of the measurement system (see Fig.~\ref{fig:02_system_overview}) is the data acquisition device, which records the current and the voltage of the aggregate load, i.e., the production machine. Since most production machines are connected via three phases, three current and three voltage signals need to be recorded. Based on these signals, the active power demand of the constituent electric loads is estimated through a combination of disaggregation algorithms. The type of the constituent loads can be predicted through labeling algorithms.

\begin{figure} [htb!]
	\centering
	\includegraphics[width=0.8\linewidth]{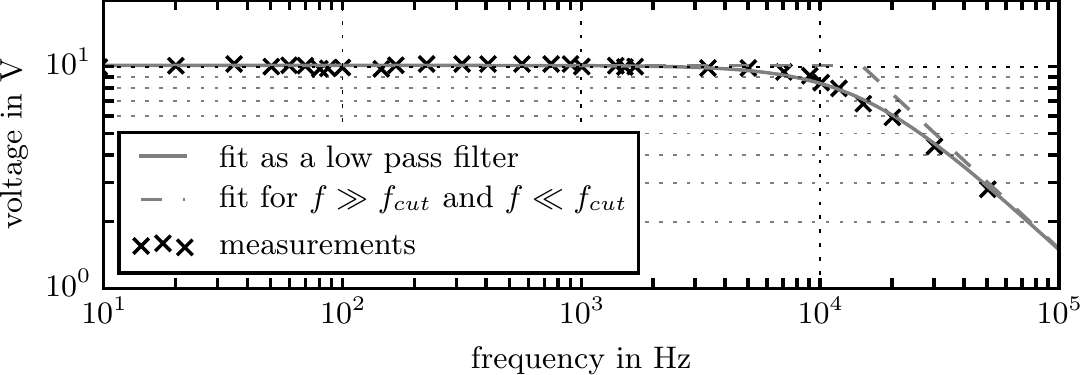}
	\caption{Experimentally determined frequency response of the utilized current clamps using an analog signal generator. A first-order low pass filter was fitted to the measured data as \mbox{$v(f)=v_0 / \sqrt{1+(f/f_{cut})^2}$}. The current clamps return a voltage signal, hence the labeling of the y-axis.}
	\label{fig:data_current_clamps}
\end{figure}

The aggregate current and voltage were measured with a sample frequency of 10 kHz using a data acquisition hardware of type DS-NET from the company DEWESoft GmbH. A V4-HV module was used for measuring the voltage, and a V8 module combined with current clamps was employed to measure the current. As current clamps, the model WZ12B from the company GMC-I Messtechnik GmbH were used, for which a measurement error of $\epsilon(t)=\pm1.5~\% \cdot i(t) \pm 1mA$ is stated. Even though they are only specified for a frequency range of 45~Hz to 500~Hz, it was experimentally determined by both the author and the manufacturer that they measure a current with a frequency of 10 kHz with an attenuation factor of only a few percent (see Fig.~\ref{fig:data_current_clamps}).

To be able to evaluate the results of the disaggregation algorithms, the current and voltage of the constituent loads were measured as well using the same data acquisition hardware and a time synchronicity of approximately 1 ms. These signals were only used for validating the disaggregation algorithms and are NOT a necessary input of the proposed measurement system.

\section{Algorithms for disaggregation}
\begin{figure} [htb!]
	\centering
	\includegraphics[width=0.8\linewidth]{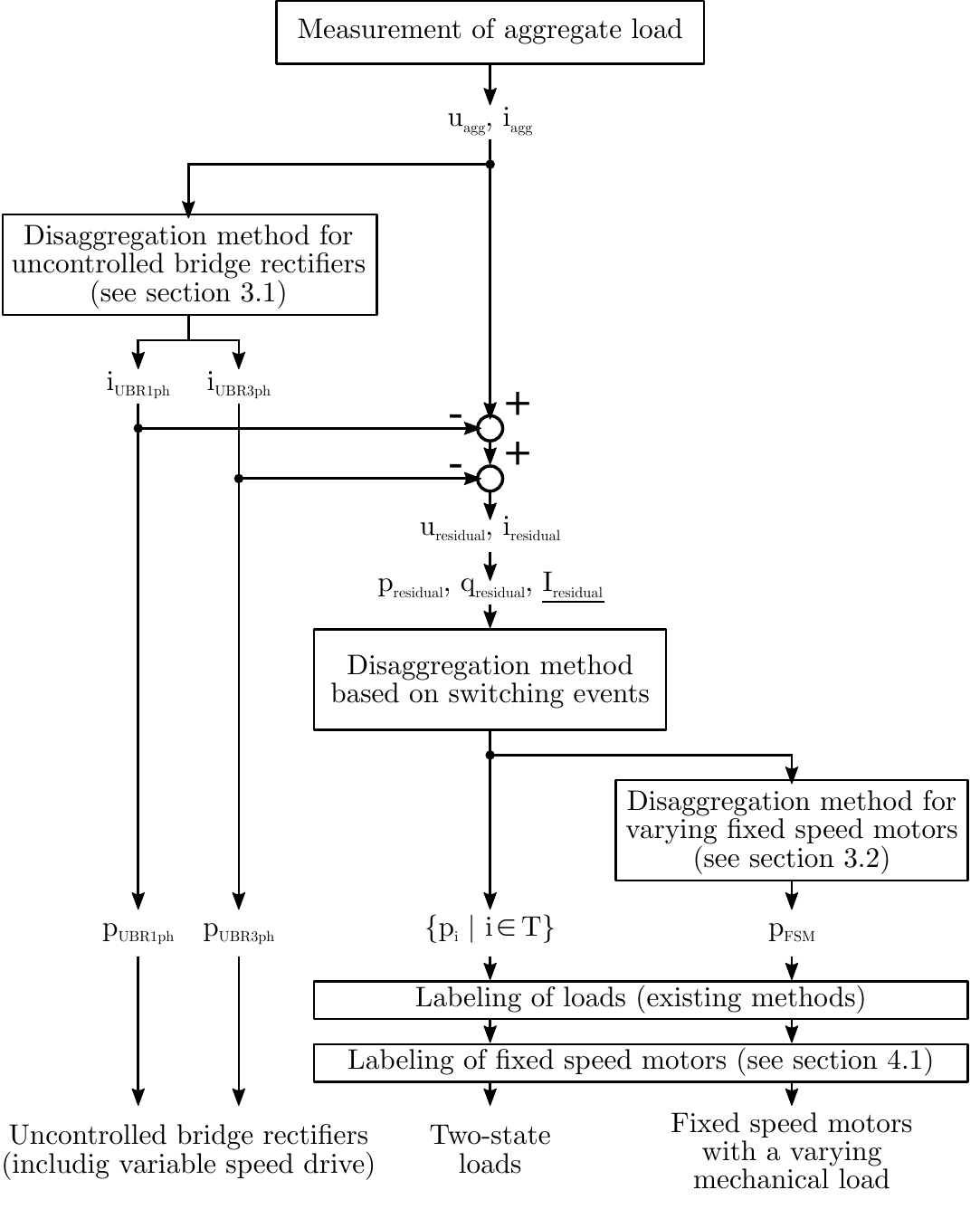}
	\caption{Flow chart of the combined measurement method based on disaggregation}
	\label{fig:03_overview_disagg}
\end{figure}

The system presented in this paper comprises three types of disaggregation methods (see Fig.~\ref{fig:03_overview_disagg}):
\begin{itemize}
	\item A disaggregation method based on switching events
	\item A new disaggregation method for uncontrolled bridge rectifiers (preliminary version published in \cite{gebbe2017nilm})
	\item  A new disaggregation method for fixed speed motors whose mechanical load varies continuously (preliminary version published in \cite{gebbe2017cost})
\end{itemize}
These methods were selected based on an analysis of 151 electric loads, which revealed that the four most frequent load classes in machines used in manufacturing were fixed-speed motors (38~\% of all loads), motors controlled by a variable speed drive (26~\%), Joule heating elements (15~\%) and rectifiers supplying different electronic loads such as programmable logic controllers (13~\%). The residual loads (8~\%) were different light sources, ultrasound generators and plasma generators.

\begin{figure} [htb!]
	\centering
	\includegraphics[width=0.8\linewidth]{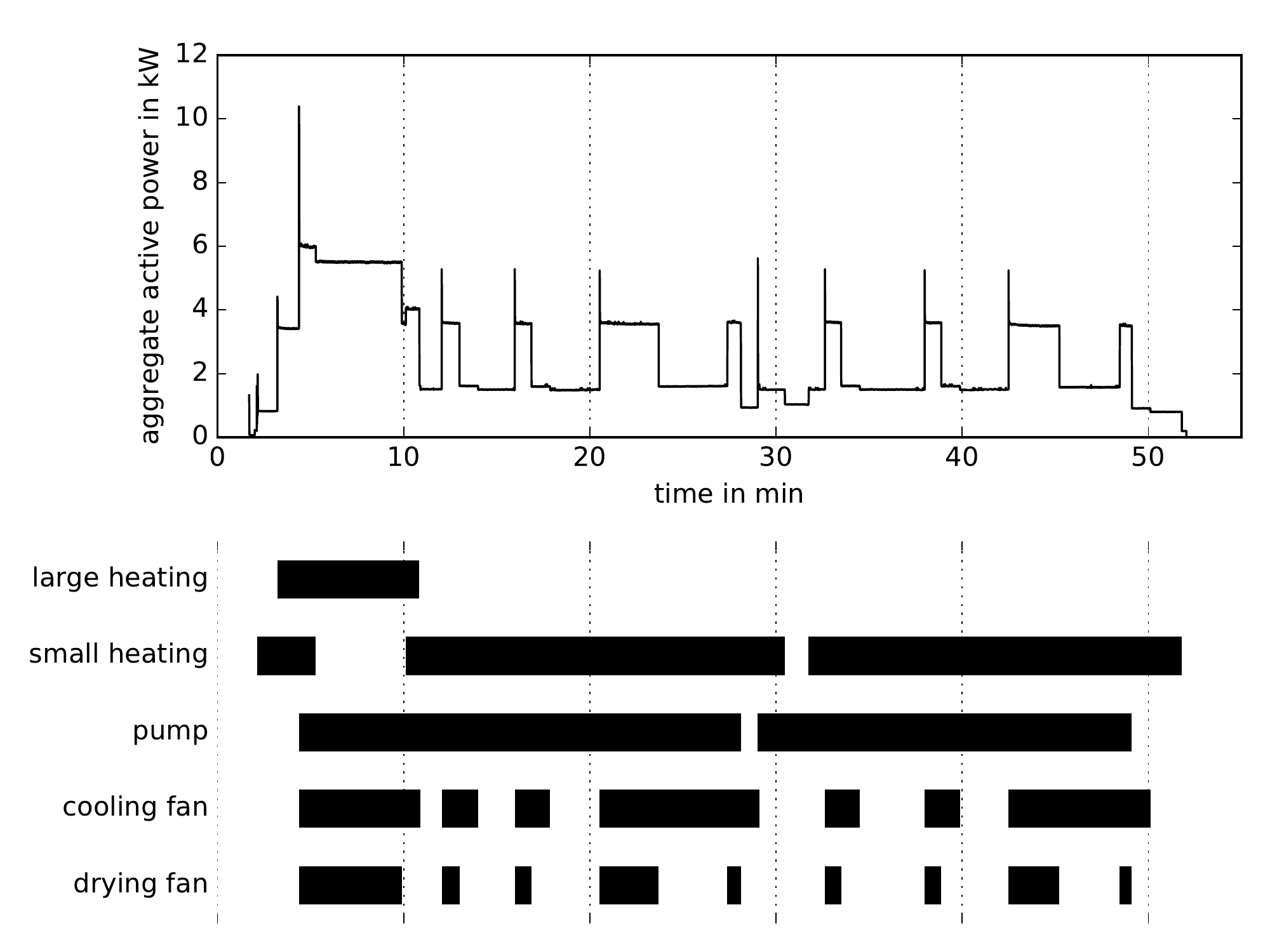}
	\caption{The aggregate active power demand of an ultrasonic cleaning system (top) is approximately constant in between switching events. Its five main constituent electric loads have only two operating states, namely on and off (on-phase is displayed in the bottom).}
	\label{fig:03_power_ultrasonic}
\end{figure}

\begin{figure} [htb!]
	\centering
	\includegraphics[width=0.8\linewidth]{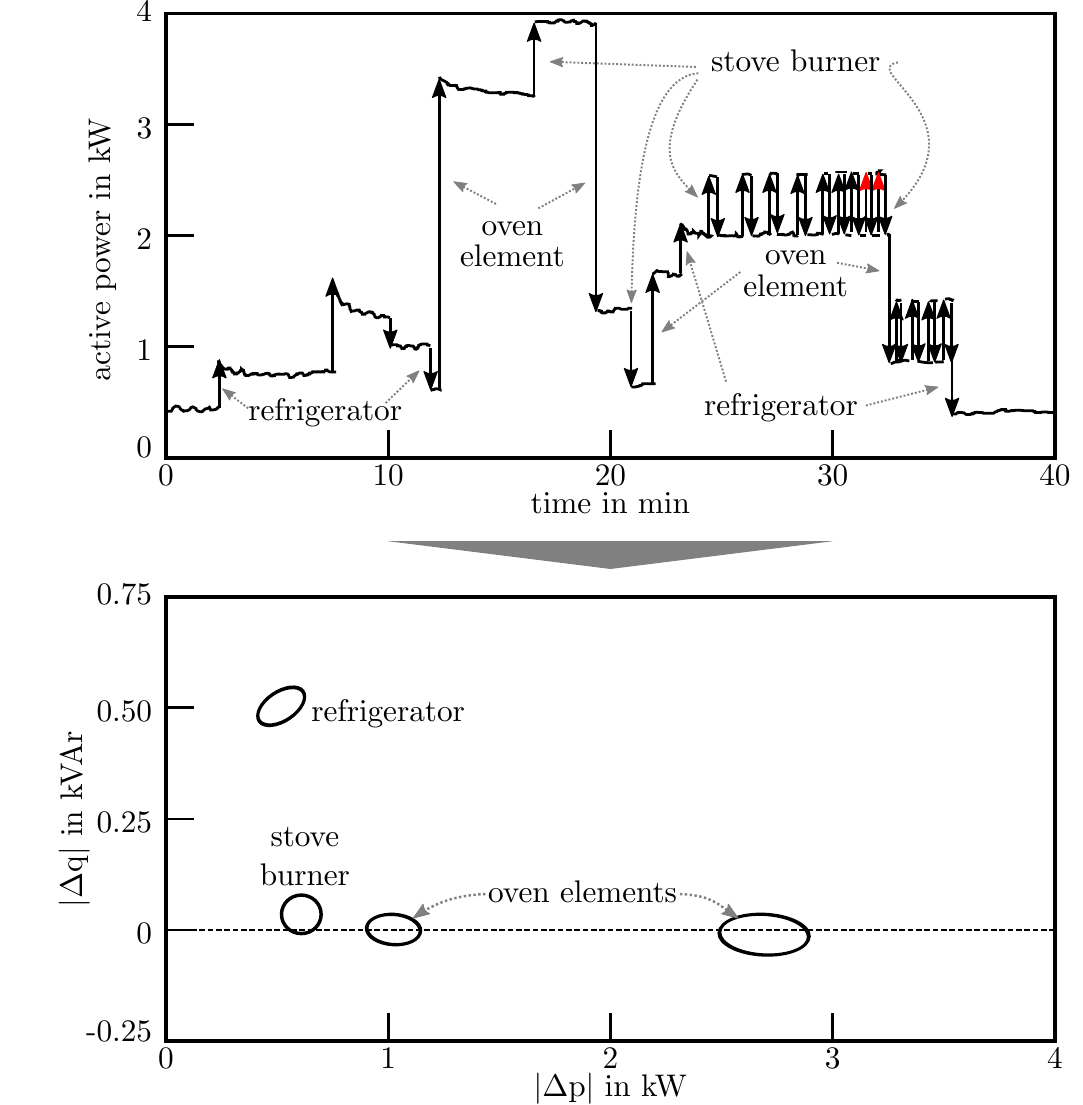}
	\caption{The disaggregation method based on switching events detects characteristic step changes in the aggregate active power demand (top) and clusters them subsequently (bottom). Based on \cite{Hart1992}.}
	\label{fig:03_Hart_combined}
\end{figure}

Most analyzed fixed-speed motors exhibited a constant energy demand when switched on, indicating a constant mechanical load (see Fig.~\ref{fig:03_power_ultrasonic}). The same behavior was observed for all Joule heating elements and some rectifiers. Such loads can be called two-state loads since their energy demand can be modeled with only two states, namely on and off. Their energy demand can be estimated using the disaggregation method based on switching events. This method was first introduced in 1985 \cite{Hart1985} and is based on the observation that each switching event of an appliance, i.e. from on to off or vice versa, leads to a characteristic step change in the active and reactive power (see Fig.~\ref{fig:03_Hart_combined} top). By clustering such detected step changes (see Fig.~\ref{fig:03_Hart_combined} bottom), the active power demand and operation schedule of individual loads can be inferred. The method was further modified in \cite{Luo2002} and adapted here with minor changes.

However, not all loads could be modeled as a two-state load: Nearly all motors controlled by a variable speed drive exhibited a continuously varying energy demand. Similarly, the energy demand of a few fixed-speed motors fluctuated significantly due to continuously varying mechanical loads. For these electric loads, the disaggregation method based on switching events is not suitable. Because of that, the two new disaggregation algorithms were developed.

All 40 variable speed drives analyzed for this paper contained uncontrolled bridge rectifiers, whose presence leads to a characteristic current shape. This fact is utilized in the new disaggregation method for uncontrolled bridge rectifiers which is explained in detail in section 3.1. Such a characteristic current shape was not found in the current draw of fixed-speed motors with a continuously varying mechanical load. Thus, another new disaggregation method was proposed for these loads, which is described in section 3.2.

Other disaggregation algorithms \cite{Zoha2012, Zeifman2011} have been considered but were dismissed for the following reasons:
\begin{itemize}
	\item Disaggregation methods using deep neural networks \cite{Kelly2015, Mauch2016, Bonfigli2018} were dismissed because they require extensive training data which cannot be easily supplied for production machines.
	\item Disaggregation methods formulated as an optimization problem \cite{Egarter2013, suzuki2008nonintrusive} were omitted because they did not offer any significant advantages over the disaggregation method based on switching events in this application scenario. Moreover, they are less anchored to the physical interpretation of the signal.
	\item Disaggregation methods based on correlation \cite{Laughman2003, Lee2005, Wichakool2009, Wichakool2015} can also be used to estimate the active power demand of uncontrolled bridge rectifiers. However, compared with the new method for uncontrolled bridge rectifiers, such methods exhibit several disadvantages:
	\begin{itemize}
		\item The correlation function needs to be manually determined for each specific variable speed drive, otherwise significant estimation errors are to be expected.
		\item The method is susceptible to other loads generating current harmonics.
		\item The method is not suitable if the aggregate load includes both a three-phase and a single-phase uncontrolled bridge rectifier.
		\item It only estimates the active power demand or the fundamental current harmonic of the uncontrolled bridge rectifier instead of its current. This is adverse if other methods work directly on the current such as the labeling method described in subsection 4.1.
	\end{itemize}
\end{itemize}

\subsection{Disaggregation algorithm for uncontrolled bridge rectifiers}
\begin{figure} [htbp!]
	\centering
	\includegraphics[width=0.8\linewidth]{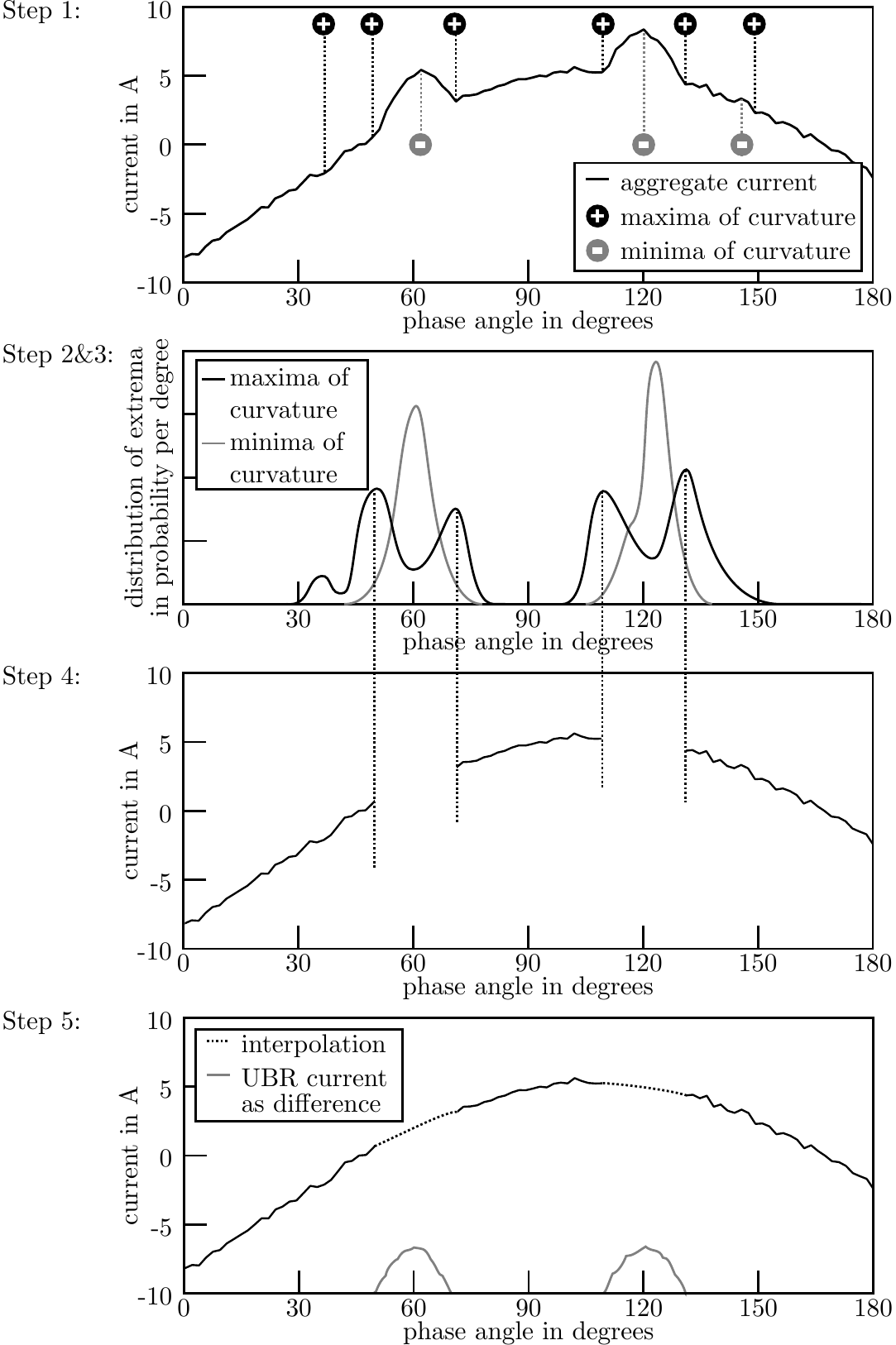}
	\caption{Disaggregation approach for uncontrolled bridge rectifiers (UBR).}
	\label{fig:06_VSD_approach}
\end{figure}

The idea for a new disaggregation method for uncontrolled bridge rectifiers (and thus also for variable speed drives featuring uncontrolled bridge rectifiers) originates from two observations: First, the characteristic peaks in the current of a three- or single-phase uncontrolled bridge rectifier are also visible in the aggregate current in most cases. Second, the current of an uncontrolled bridge rectifier is zero for every region apart from the peaks. Hence, if the visible peaks in the aggregate current could be filtered or 'cut off', the aggregate current could be separated into the current of the uncontrolled bridge rectifier and the current of all residual loads. This approach can be realized by performing the following steps (see Fig.~\ref{fig:06_VSD_approach}):
\begin{enumerate}
	\item Detect the beginning and the ending of peaks by analyzing the curvature of the aggregate current.
	\item Identify the type of uncontrolled bridge rectifier, i. e. either single-phase or three-phase, by examining the distribution of all estimated peak beginnings and endings.
	\item Filter out false positives and false negatives in the estimated current peak beginnings and endings.
	\item Estimate the current of the residual loads through interpolation.
	\item Estimate the current of the uncontrolled bridge rectifier through a simple subtraction.
\end{enumerate}

This method does not exhibit any of the disadvantages of the disaggregation method based on correlation:

First, the method yields accurate results for a variety of uncontrolled bridge rectifiers without requiring any training data. In order to show this, an aggregate load was simulated comprising an uncontrolled bridge rectifier and a linear load which can be described with a pure sinus whose peak current amplitude was twice as high as the peak current amplitude of the uncontrolled bridge rectifier. The active power demand of the uncontrolled bridge rectifier was then estimated using the new disaggregation method with an accuracy of 95-100~\% depending on the specific electric structure of the uncontrolled bridge rectifier. The accuracy was calculated as

\begin{align}
\label{eq:accuarcy}
acc &=1- \frac{\Delta E}{E}
\\ \nonumber
&=1- \frac{\int|p_{est} (t)-p_{true} (t)|dt}{\int |p_{true} (t)|dt}
\end{align}

, where $p_{est}$ refers to the estimated power using the disaggregation method and $p_{true}$ to the true power determined by measuring the current and voltage of the electric load directly.

\begin{figure} [htb!]
	\centering
	\includegraphics[width=0.8\linewidth]{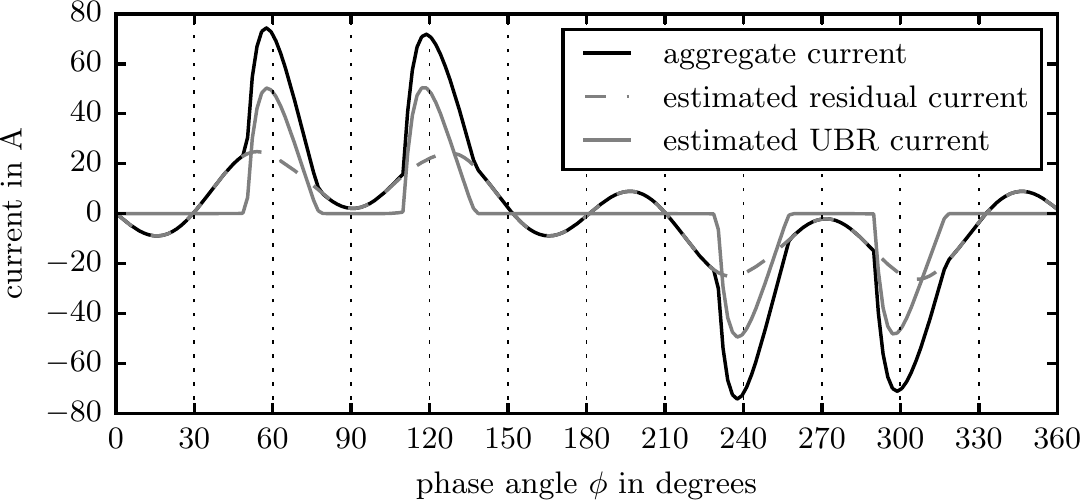}
	\caption{Disaggregation results when another load with current harmonics is present in the aggregate load}
	\label{fig:06_VSD_robustness}
\end{figure}

Second, the method is less susceptible to other loads generating current harmonics. In order to show this, again an aggregate load comprising an uncontrolled bridge rectifier plus another load whose first and fifth current harmonic equal the ones of the rectifier in magnitude and phase was simulated. Both loads could be disaggregate with accuracies above 97~\% (see Fig.~\ref{fig:06_VSD_robustness}). In contrast to that, the disaggregation method based on correlation would only yield an accuracy of approximately 0~\% for the uncontrolled bridge rectifier.

Third, the method is also suitable if the aggregate load contains both a three-phase and a single-phase uncontrolled bridge rectifier (see Fig. 9 top). To this end, the current of such an aggregate load was simulated and the currents of both uncontrolled bridge rectifiers were estimated using the new disaggregation method. Thereby, accuracies of 98~\% and 99~\% were achieved.

\begin{figure} [htb!]
	\centering
	\includegraphics[width=0.8\linewidth]{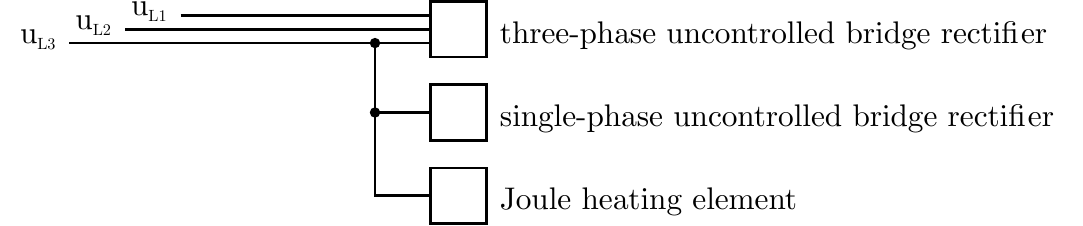}
	\includegraphics[width=0.8\linewidth]{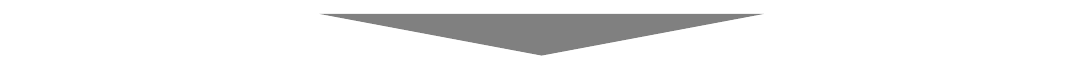}
	\includegraphics[width=0.8\linewidth]{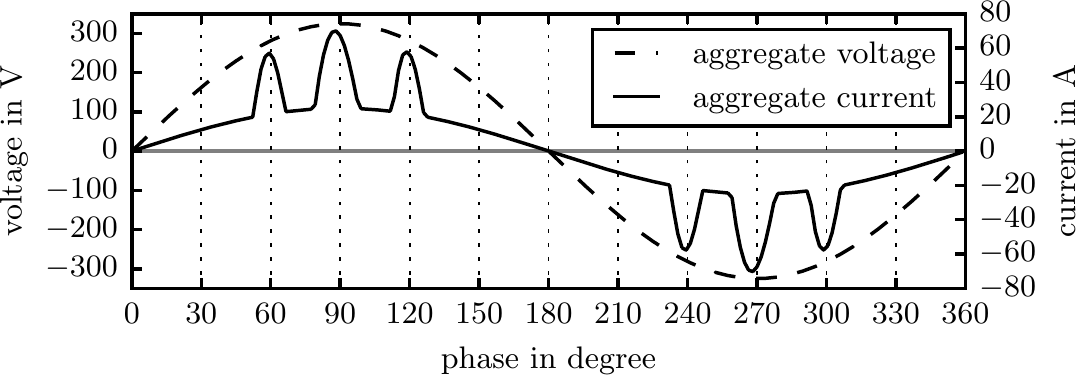}
	\includegraphics[width=0.8\linewidth]{img/05_arrow}
	\includegraphics[width=0.8\linewidth]{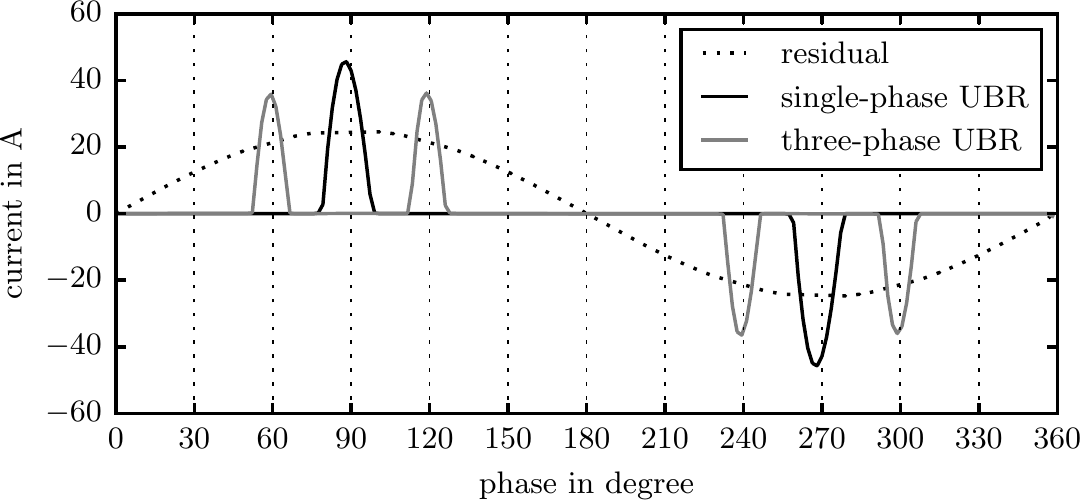}
	\caption{If both a three-phase and a single-phase rectifier are present in an aggregate load (top), the aggregate current (mid) can be automatically disaggregated into its constituent parts (bottom).}
	\label{fig:06_VSD_robustness_both}
\end{figure}

Fourth, the method works directly on the current (many samples per period) instead of the active power demand (at most one sample per period) and thus preserves more information.

\subsection{Disaggregation algorithm for fixed speed motors whose mechanical load varies continuously}

An algorithm was developed for a situation in which the following conditions hold true:
\begin{itemize}
	\item The aggregate load consists of only one fixed speed motor whose active power demand varies continuously, and additionally any number of two-state loads (including fixed speed motors with a constant mechanical load) and uncontrolled 
	bridge rectifiers.
	\item The fixed speed motor with a continuously variable power demand exhibits a characteristic step change in the active power demand and in the current harmonics when switched on and off.
	\item There is a significant linear correlation between at least one current harmonic of the aggregate load and the continuously variable active power demand of the fixed speed motor.
\end{itemize}

\begin{figure} [htbp!]
	\centering
	\includegraphics[width=0.8\linewidth]{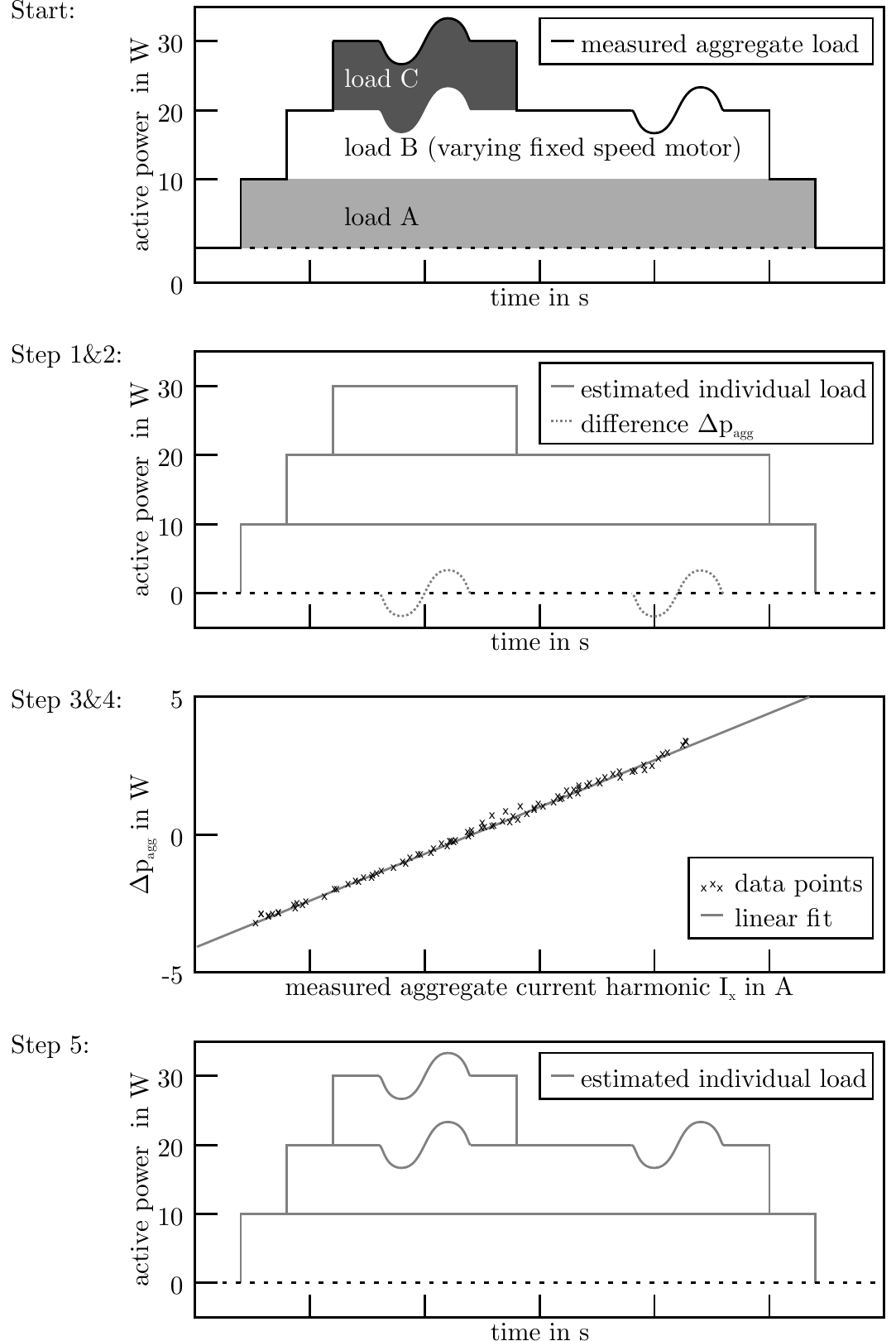}
	\caption{Disaggregation approach for fixed speed motors whose mechanical load varies continuously}
	\label{fig:06_FSM_approach}
\end{figure}

In this case, the continuously variable active power demand of the fixed speed motor can be estimated by performing the following steps (see Fig.~\ref{fig:06_FSM_approach}):

\begin{enumerate}
	\item Perform the existing disaggregation method for uncontrolled bridge rectifiers as well as the disaggregation method based on switching events (see flow chart in Fig. 4).
	\item Calculate the difference $\Delta p_{agg}$ between the sum of the estimated active power demand of the loads and the measured aggregate active power.
	\item Identify a specific current harmonic $I_x$, which correlates linearly with $\Delta p_{agg}$ and identify to which load it belongs to.
	\item Perform a linear fit between $\Delta p_{agg}$ and $I_x$.
	\item Estimate the active power demand of the fixed speed motor through the linear fit.
\end{enumerate}

The algorithm is based on the following idea: In order to derive a continuously variable active power demand of a specific load based on only measurements of the aggregate load, it is necessary that one aggregate feature can be uniquely associated with the load. To this end, only two electrical properties have been proposed in literature so far, namely high frequency voltage noise \cite{Gupta2010, Patel2007} and current harmonics \cite{Laughman2003, Lee2005, Wichakool2009, Wichakool2015}. High frequency harmonics could not be analyzed in this thesis, because it requires expensive measurement equipment featuring a sample frequency of 1 MHz which was not available. Current harmonics have already been used for the extraction of the active power demand of uncontrolled bridge rectifiers \cite{Laughman2003, Lee2005, Wichakool2009, Wichakool2015}, which are characterized by a high total harmonic distortion of more than 100~\%. In contrast to that, the current of fixed speed motors typically exhibits a total harmonic distortion of only a few percent. Moreover, there is no obvious relation between any current harmonic and the active power demand. Nevertheless, it was found that a significant linear correlation between a current harmonic and the active power demand can exist.

An interesting question is when and why such a linear correlation exists. While the relation between current harmonics and the fundamental current harmonic (with which the active power demand can be calculated) can be analytically derived for several electric loads such as uncontrolled bridge rectifiers, light dimmers and ideal line commuted converters \cite{grady2012understanding}, it cannot be derived in a similar way for fixed speed motors. In \cite{grady2012understanding}, it is noted that single-phase motors typically have a low total harmonic distortion of around 10~\% , because the current becomes nonlinear when the motor is operated with peak flux densities beyond the saturation knee. No analytic relation is specified, though. Moreover, almost all ideal electric loads in AC systems generate only odd harmonics, since loads typically affect the positive and negative cycles symmetrically \cite{chapman2005electric}. However, in the validation example shown below the linear relation between an even current harmonic and the active power demand is used. Reasons for even harmonics are either loads such as diodes or imperfections of AC loads such as tolerances in transformer windings, commutation reactances or deviations in the firing times of thyristors \cite{Buddingh2003a}. However, in general, even harmonics seem to be scarcely described in literature \cite{Barros2007} and it is difficult to specify the conditions under which a linear correlation between the active power demand and an even current harmonic exists.

The new disaggregation method was used to extract the active power demand of a fixed speed motor driving a vacuum pump which varied between 600~W and 950~W (see Fig. 11). The motor is part of a thermoform machine which also comprises several Joule heating elements, two variable speed drives, three fixed speed motors driving different types of conveyor belts, a fan and an external cooling unit consisting of a compressor, a pump and a fan. Despite this challenging scenario the active power demand of the fixed speed motor driving the vacuum pump could be extracted with an accuracy of 89~\%.

\begin{figure} [htb!]
	\centering
	\includegraphics[width=0.8\linewidth]{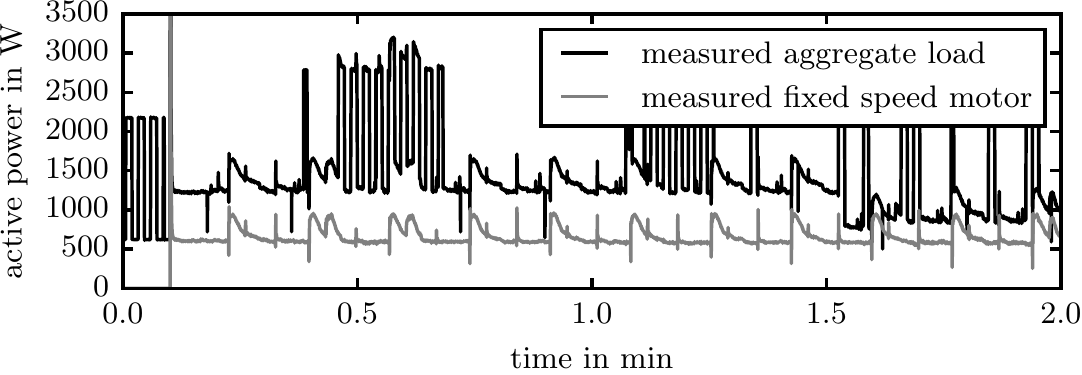}
	\includegraphics[width=0.8\linewidth]{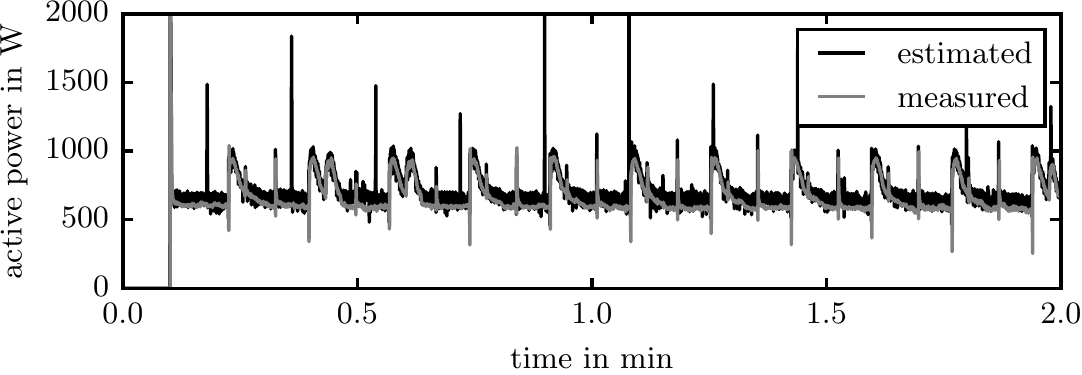}
	\caption{The measured active power demand of the aggregate load and of the fixed speed motor (top). Despite its continuous variation, it is estimated with an accuracy of 89~\% (bottom).}
	\label{fig:06_FSM_valid_step0}
\end{figure}

\section{Algorithms for labeling}

The disaggregation methods presented in section~3 yield the estimated active power demand for all identified loads. Moreover, they intrinsically specify the type of load: For example, all loads whose active power demand was estimated using the disaggregation algorithm for uncontrolled bridge rectifiers are assumed to be uncontrolled bridge rectifiers (the algorithm even guesses whether it is a single- or three-phase bridge rectifier). Similarly, the disaggregation algorithm for fixed speed motors whose mechanical load varies continuously is assumed to only yield the active power demand for these kinds of loads. This is also true for the disaggregation method based on switching events, which targets two-state loads. However, two-state loads comprise a large variety of load types ranging from Joule heating elements to fixed speed motors and light sources. Thus, it would be beneficial for the end-user if the system could provide a more specific load label.

One approach to specify the type of the load is by analyzing its power factor $λ = p/s$. In this way, fixed speed motors can be differentiated from Joule heating elements or capacitve elements, because their power factors are signifcantly different. This approach has already been used since the 1990 \cite{Hart1992} and is not analyzed here further.

An additional approach was tested here: Since fixed speed motors represent the most frequent load class in this paper (see section 3), it was tested, whether they can be further distinguished according to the type of mechanical load they drive, i.e. a compressor, a fan, a pump or another mechanical load. This analysis was triggered by \cite{Sultanem1991}, in which it is indicated that a differentiation of motors according to their mechanical load is feasible using the transient turn-on current. Moreover, the transient behavior of a typical load is known to be intimately related to the physical task that the load performs \cite{Leeb1993}. However, whether such a differentiation is actually possible has never been tested, nor has a specific labeling method been proposed. Therefore, the following analysis was carried out.

\subsection{Analysis of classifying motors according to their mechanical load}

\begin{figure} [htb!]
	\centering
	\includegraphics[width=0.8\linewidth]{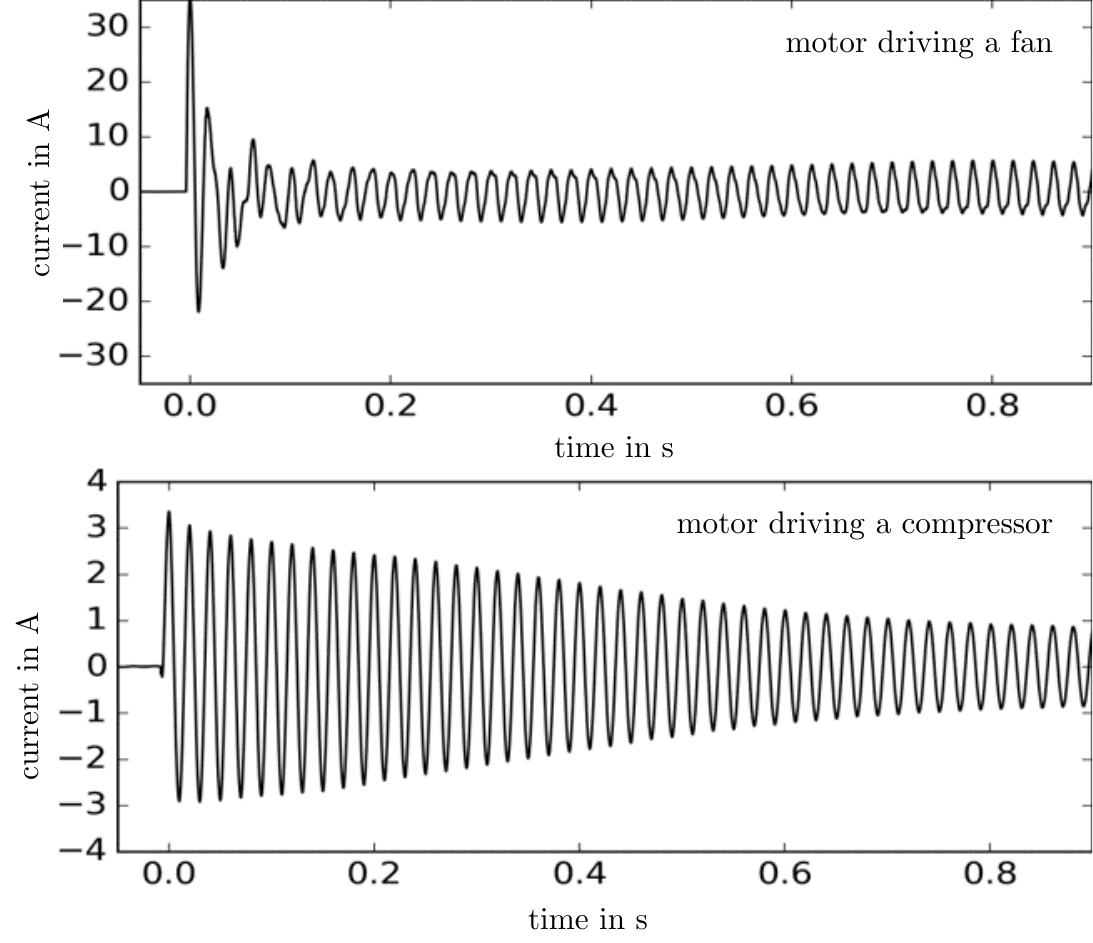}
	\caption{The turn-on transient current of two motors, one driving a fan (top), the other one driving a compressor (bottom).}
	\label{fig:example_transient_current}
\end{figure}

A manual analysis of some transient turn-on currents strenghtens the hypothesis that this electrical property could be used to automatically distinguish fixed speed motors driving e.g. fans from fixed speed motors driving compressors or pumps (see Fig.~\ref{fig:example_transient_current}). 

In order to strenghten or weaken this hypothesis, we conducted an empirical study in which we recorded the current (and voltage) of 18 different fixed-speed motors (see Table~\ref{tab:04_data_motors}). From this recorded transient turn-on current more than 100 different features were extracted (e.g. exponential decay constant, current peak values, total harmonic distortions) and a classifier was trained in a supervised manner using different combinations of these features. An important step was to normalize the current for each load initially such that the steady-state current amplitude was 1~A. Therefore, the only difference between the currents was their transient behavior.

\begin{table}[htb!]
	\caption{Overview of analyzed fixed-speed motors (SLM=Selective laser sintering machine)}
	\begin{tabularx}{\textwidth}{Xlllp{2cm}p{2cm}}
		\toprule
		Machine & Pump & Compressor & Fan & Other mech. Load & Number of turn-on events \\
		\midrule
		Thermoform machine - vacuum pump &  & 1 &  &  & 14 \\ 
		Thermoform machine - conveyor belt &  &  &  & 1 & 12 \\ 
		Thermoform machine - motor for winding &  &  &  & 1 & 21 \\ 
		Thermoform machine - cooling unit & 1 &  &  &  & 12 \\ 
		Thermoform machine - cooling unit &  & 1 &  &  & 8 \\ 
		Thermoform machine - cooling unit &  &  & 1 &  & 9 \\ 
		Heated washing basin & 1 &  &  &  & 27 \\ 
		Milling machine - cooling lubricant pump & 1 &  &  &  & 8 \\ 
		Motor driving a generator &  &  &  & 1 & 10 \\ 
		Pedestal fan &  &  & 1 &  & 28 \\ 
		Cooling unit for SLM 1 & 1 &  &  &  & 15 \\ 
		Cooling unit for SLM 1 &  & 1 &  &  & 38 \\ 
		Cooling unit for SLM 1 &  &  & 1 &  & 39 \\ 
		Cooling unit for SLM 2 & 1 &  &  &  & 47 \\ 
		Cooling unit for SLM 2 &  & 1 &  &  & 46 \\ 
		Ultrasonic cleaner - fan &  &  & 1 &  & 14 \\ 
		Ultrasonic cleaner - fan for drying &  &  & 1 &  & 18 \\ 
		Ultrasonic cleaner - circulation pump & 1 &  &  &  & 10 \\ 
		\midrule
		SUM & 6 & 4 & 5 & 3 & 376 \\
		\bottomrule
	\end{tabularx}
	\label{tab:04_data_motors}
\end{table}

We will conclude the results of this empirical study by stating that we did NOT find a classifier which performed significantly better than a random classifier in distinguishing fixed speed motors according to their type of mechanical load. While this empirical analysis does not disprove the hypothesis due to the limited data it definitely weakens it.

\begin{figure} [htbp!]
	\centering
	\includegraphics[width=0.9\linewidth]{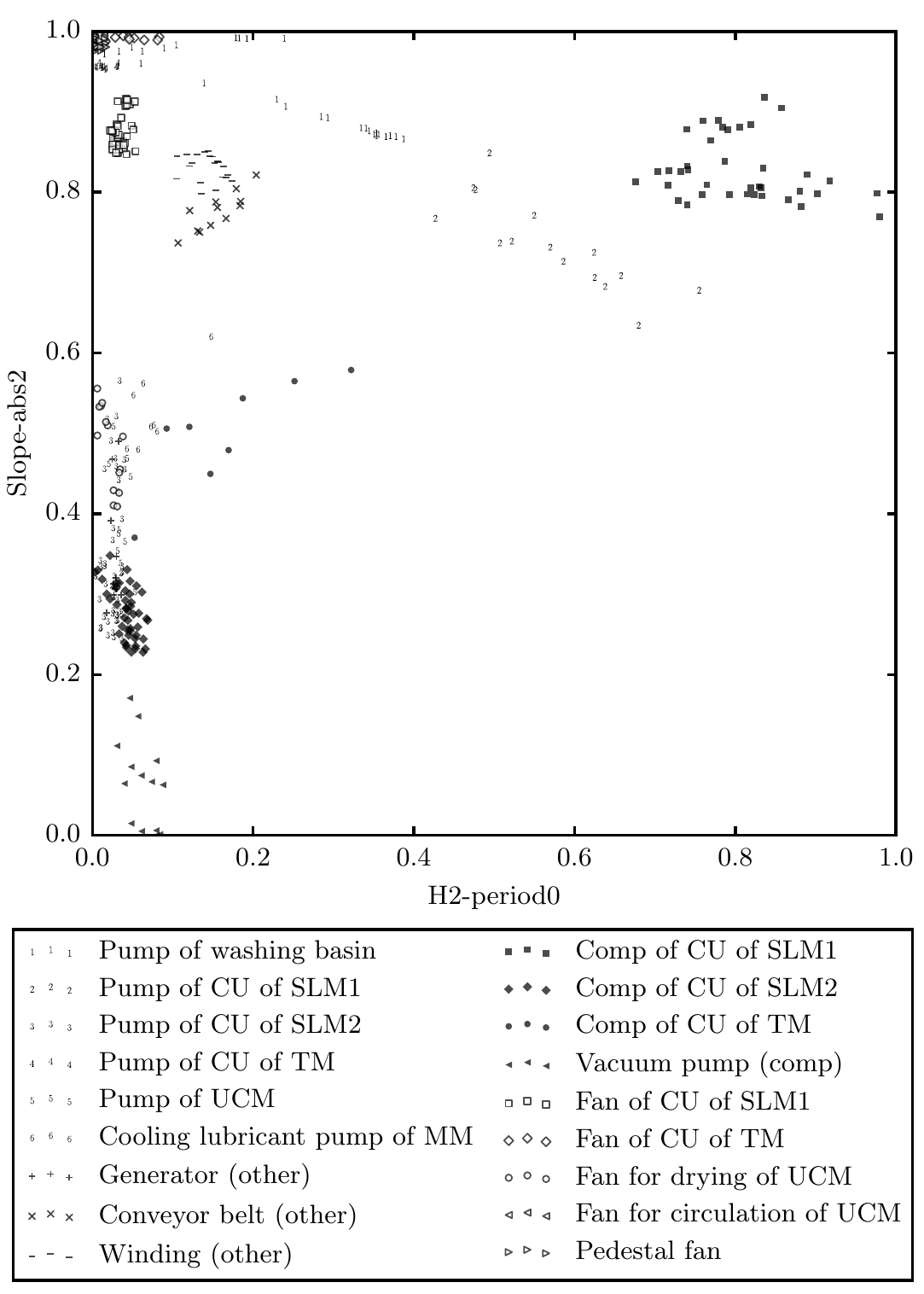}
	\caption{Scatter plot of all turn-on events. Please note, that the classifier achieving an f1-score of 98~\% did not only use the two features presented here, but twelve more features (14 in total). CU= cooling unit, SLM = selective laser sintering machine, UCM = ultrasonic cleaning machine, MM = milling machine.}
	\label{fig:04_result_motor_scatter}
\end{figure}

However, we DID find a classifer which could distinguish between each of the 18 fixed speed motors with an f1-score of 98~\%, see Fig.~\ref{fig:04_result_motor_scatter} (in contrast, a random classifier would have an f1-score of approximately $1/18=5.6~\%$). This means that each load has a rather unique fingerprint in the transient turn-on current. However, this unique fingerprint seems to vary signifcantly among fixed speed motors driving the same type of mechanical load, so that a better than random classification is not possible using only the transient turn-on current.

For a much more detailed report we refer to the publicly accessible paper on \cite{gebbe2019supervised}.

\newpage
\section{Implementation}

The algorithms for disaggregation and labeling were implemented in python and were applied on the recorded data afterwards, i.e. not during measurement. They could in principle also run in real-time during the measurement, but that would require complex modifications, see \cite{Hart1985}. The signals were not preprocessed, e.g. smoothed, in any way.

The measurement system was applied to several aggregate loads representing different kinds of machines used in manufacturing. Two of those results are presented in the following. In both cases the accuracy was determined as in equation~\ref{eq:accuarcy}. While the machines consume power on three phases, only one phase is considered in the following due to a limited number of current clamps for validating the disaggregation results as well as limited space in the cabinet box. However, it is straightforward to apply the algorithms to all three phases and similar results can be expected.

\begin{figure} [htbp!]
	\centering
	\includegraphics[width=0.8\linewidth]{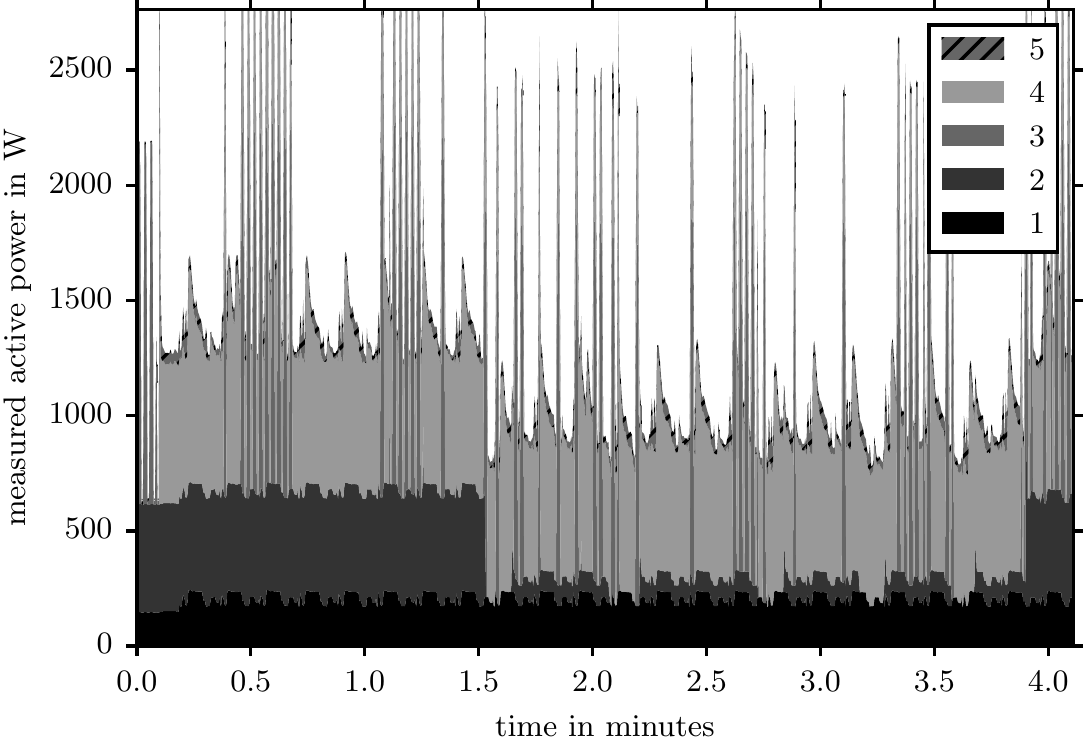}
	\includegraphics[width=0.8\linewidth]{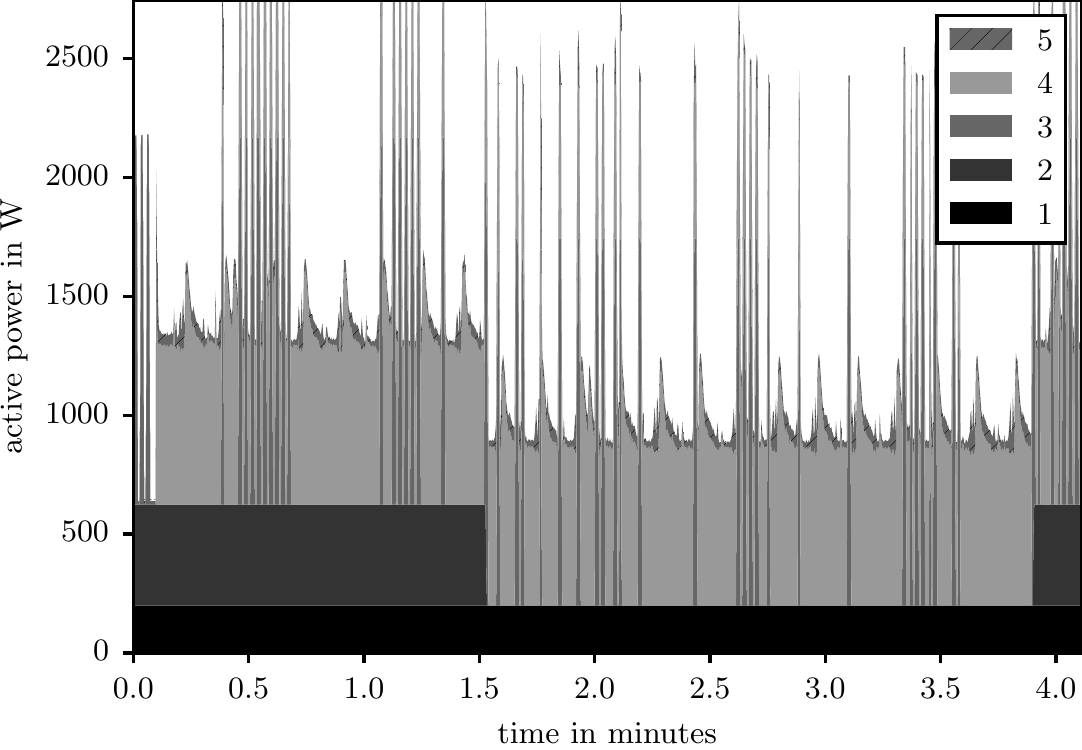}
	\vspace{0.2cm}
	\begin{tabularx}{\textwidth}{lXrrr}
		\toprule
		Label & Load  & e in Wh & $\Delta$e in Wh & acc \\
		\midrule
		1     & Rest: Two motors for winding, two fans, DC supply, conveyor belt & 14.2  & 1.7   & 88~\% \\
		2     & External cooling device & 16.3  & 4.0   & 76~\% \\
		3     & Heating & 17.3  & 1.2   & 93~\% \\
		4     & Motor driving a vacuum pump & 43.5  & 4.7   & 89~\% \\
		5     & Two variable speed drives (UBR3ph) & 4.1   & 0.1   & 97~\% \\          
		& \textbf{Sum}   & \textbf{95.4}  & \textbf{11.7}  & \textbf{88\%} \\
		\bottomrule
	\end{tabularx}%
	\caption{Measured (top) and estimated (middle) power demand of a thermoform machine}
	\label{fig:05_TM}
\end{figure}

The first example represents a thermoform machine used for packaging e.g. cold cuts (see Fig.~\ref{fig:05_TM}). By measuring only the current and voltage of the complete machine, the five most important loads could be identified. They include a fixed speed motor driving a vacuum pump, a heating element, an external cooling device and two variable speed drives featuring a three-phase uncontrolled bridge rectifier. These loads contributed more than 80~\% to the electric energy demand of the machine. Five small fixed speed motors and a DC supply were not recognized as individual loads but instead as one aggregate one. The active power demand of the identified loads was estimated with an average weighted accuracy of 88~\%. Only the external cooling device was estimated with an accuracy of less than 88~\%, namely with 76~\%, because four turn-on and turn-off events with a magnitude of only 80 W were misclassified as noise. Such a misclassification can be prevented in the future by employing a more intelligent switching event clustering algorithm. In the current implementation the events are clustered based on only the magnitude of the step change without taking into account the sequence of on and off step changes or the measured aggregate active power.

\begin{figure} [htbp!]
	\centering
	\includegraphics[width=0.8\linewidth]{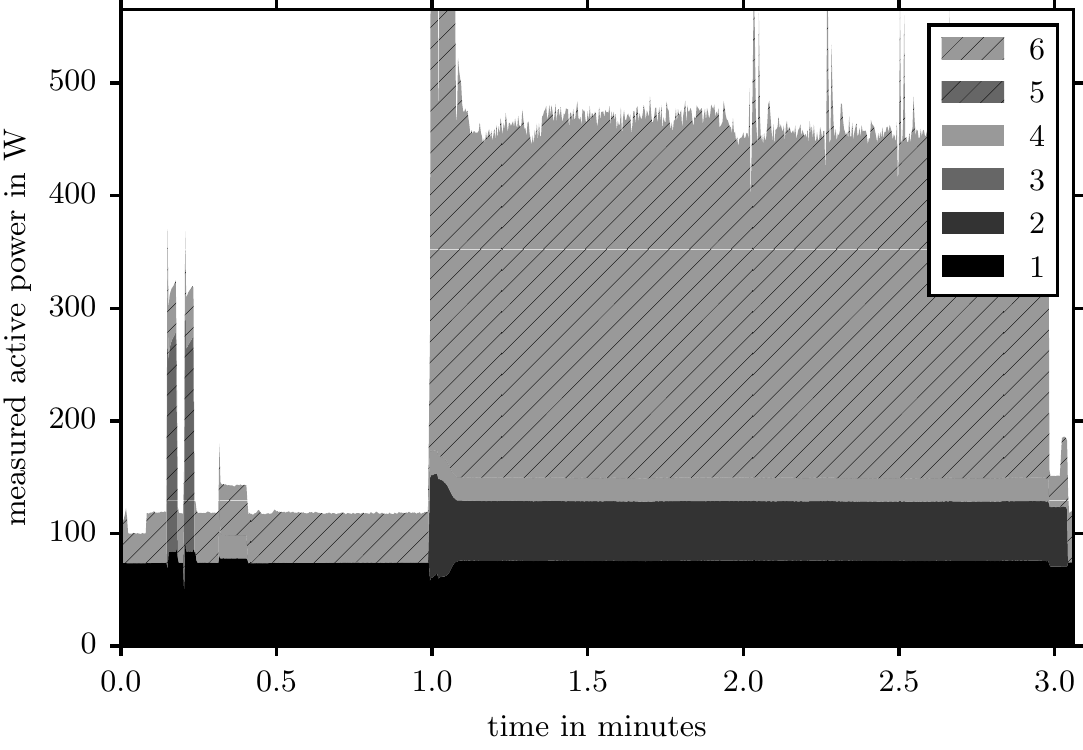}
	\includegraphics[width=0.8\linewidth]{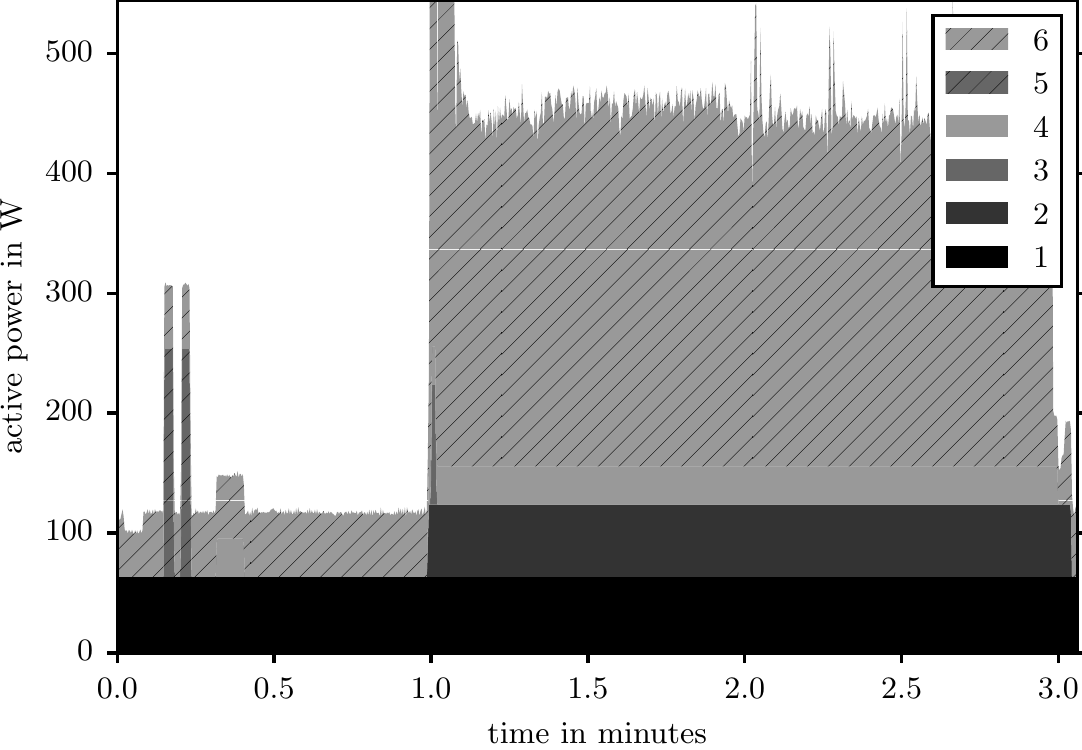}
	
	\vspace{0.2cm}
	\begin{tabularx}{\textwidth}{lXrrr}
		\bottomrule
		Label & Description of load & e in Wh & $\Delta e$ in Wh & acc \\
		\midrule
		1     & Residual loads: pump for lubricants and electronics & 2,62  & 0,63  & 76\% \\
		2     & Fan for spindle & 1.85  & 0.27  & 85\% \\
		3     & \textit{False positive} & 0.00  & 0.02  & $\infty$ \\
		4     & Pump for cooling lubricant & 0.72  & 0.40  & 45\% \\
		5     & Pump for hydraulic & 0.21  & 0.03  & 87\% \\
		6     & Variable speed drive (UBR3ph) & 11.53 & 0.83  & 93\% \\
		& \textbf{Sum}   & \textbf{16.9}  & \textbf{2.2}   & \textbf{87\%} \\
		\bottomrule
	\end{tabularx}%
	\caption{Measured (top) and estimated (middle) power demand of a milling machine}
	\label{fig:05_MM}
\end{figure}

The second example represents a milling machine consisting of a variable speed drive for the three axes and the spindle as well as several pumps, a spindle fan and electronics (see Fig.~\ref{fig:05_MM}). The disaggregation results are very similar to the ones of the thermoform machine in the sense that the identified loads contribute more than 80~\% to the total electric energy demand and that the average weighted accuracy is 87~\%. The variable speed drive, the hydraulic pump and the spindle fan were extracted with accuracies of 85~\% and more. Despite this suitable result, the low estimation error of 7~\% for the variable speed resulted in an overestimation of the active power demand of the pump for the cooling lubricant which led to an accuracy of only 45~\%. The imperfect disaggregation of the variable speed drive also yielded a false positive with a brief active state at around 1:00 min.

\newpage
\section{Summary}
This paper presented a new measurement system based on nonintrusive load monitoring for determining the electric energy demand of production machine components. It only measures the voltage and the current of the aggregate load and then performs automatic disaggregation and labeling algorithms to estimate the energy demand of the constituent loads.

While similar methods have already been used for a long time to determine the electric energy demand of household appliances, this paper represents the first instance in which such methods are used to determine the energy demand of production machine components. Moreover, a new disaggregation algorithm for variable speed drives featuring uncontrolled bridge rectifiers as well as a new disaggregation algorithm for fixed speed motors with a varying mechanical load were developed. In addition to that a new labeling algorithm was tested to differentiate fixed speed motors according to their mechanical load using the turn-on transient current.

The new measurement system was then applied to a thermoform machine and a milling machine. For those two machines, the identified loads contributed more than 80~\% to the total electric energy demand, respecitvely. Moreover, the active power demand was estimated with an average weighted accuracy of 87-88~\%. These values indicate that nonintrusive load monitoring can be successfully applied to determine the active power demand of the most relevant electric loads in a production machine with a suitable accuracy.

There are several options for further improvement: First, many production machines draw current on three phases so that the correlations between the three phases can be used. Here, each phase was treated individually. Second, the operating status of loads can be monitored additionally as indicated in \cite{Panten2016, gebbe2014estimating}. Third, the voltage can be measured with a sampling rate of 1~MHz so that the voltage noise can be used as an additional feature \cite{Gupta2010, Patel2007}. Fourth, it would be of great value to the user, if the algorithms state their confidence in their estimation.

\section{Funding}
The author would like to express their sincere thanks to the Freistaat Bayern for funding the project ”Green Factory Bavaria” in the framework of the future initiative “Aufbruch Bayern”.

\bibliographystyle{unsrt} 
\bibliography{library,library_else,library_CIRP}

\end{document}